\documentclass[12pt]{article}
\usepackage{amssymb}
\usepackage{bbm}
\usepackage{epsfig}
\usepackage{array}
\usepackage{float}
\usepackage{dsfont}
\usepackage{amstext}
\usepackage{amsmath}
\usepackage{psfrag}
\usepackage{a4}
\usepackage{a4wide}
\usepackage{hyperref}
\usepackage{feynmp-auto}

\usepackage{graphicx}
\usepackage{subcaption}
\usepackage{epsfig}
\usepackage{epstopdf}
\usepackage{xcolor}

\usepackage[T1]{fontenc}

\DeclareSymbolFont{myletters}{OML}{ztmcm}{m}{it}
\DeclareMathSymbol{\uplambda}{\mathord}{myletters}{"15}
\DeclareMathSymbol{\upxi}{\mathord}{myletters}{"18}

\usepackage{amsfonts}

\usepackage{lmodern} 
\usepackage{amsmath}                                                    
\usepackage{amsthm}                                                     
\usepackage{amssymb}
\usepackage{multirow}
\numberwithin{equation}{section} 
\newcommand{\newc}{\newcommand}
\newc{\be}{\begin{equation}}
\newc{\ee}{\end{equation}}
\newc{\bg}{\begin{gathered}}
\newc{\eg}{\end{gathered}}
\newc{\tref}[1]{Table \ref{#1}}
\newc{\eref}[1]{Equation \eqref{#1}}
\newc{\su}[1]{$SU(#1)$}
\newc{\bm}[1]{\mathbf{#1}}
\newc{\fref}[1]{Figure \ref{#1}}

\newc{\ra}{\rightarrow}
\newc{\lra}{\leftrightarrow}
\newc{\ov}{\overline}
\newc{\ba}{\begin{eqnarray}}
\newc{\ea}{\end{eqnarray}}
\newc{\mf}{\mathsf}

\usepackage{hyperref}

\def\be{\begin{equation}}
\def\ee{\end{equation}}
\def\bea{\begin{eqnarray}}
\def\eea{\end{eqnarray}}

\begin{document}

\begin{center}
\baselineskip 20pt 
{\Large\bf Constraining the Cosmic Strings Gravitational Wave Spectra in No Scale Inflation with Viable Gravitino Dark matter and Non Thermal Leptogenesis}

\vspace{1cm}

{\large 
	Waqas Ahmed$^{a}$ \footnote{E-mail: \texttt{\href{mailto:waqasmit@hbpu.edu.cn}{waqasmit@hbpu.edu.cn}}},
		M. Junaid$^{b}$ \footnote{E-mail: \texttt{\href{mailto:umer@udel.edu}{mjunaid@ualberta.ca}}},
			Salah Nasri$^{c, d}$  \footnote{E-mail: \texttt{\href{mailto:snasri@uaeu.ac.ae; salah.nasri@cern.ch}{snasri@uaeu.ac.ae; salah.nasri@cern.ch}}} and Umer
	Zubair$^{e, f}$ \footnote{E-mail: \texttt{\href{mailto:umer@udel.edu}{umer@udel.edu}}} 
	
} 
\vspace{.5cm}

{\baselineskip 20pt \it
	$^{a}$ \it
	School of Mathematics and Physics, Hubei Polytechnic University, \\
	Huangshi 435003,
	China \\
	\vspace*{6pt}

	$^{b}$National Centre for Physics, Islamabad, Pakistan \\
	\vspace*{6pt}
		$^{c}$Department of Physics, United Arab Emirates University,\\
		Al Ain 15551 Abu Dhabi, UAE \\
		\vspace*{6pt}
		$^{d}$International Center for Theoretical Physics, Trieste, Italy\\
		\vspace*{6pt}
	$^e$Department of Physics and Astronomy, University of Delaware,\\ Newark, DE 19716, USA \\
	\vspace*{6pt}
	$^f$College of Humanities and Sciences, Thomas Jefferson University,   \\
	East Falls Campus, Philadelphia, PA 19144, USA \\
	
	\vspace{2mm} }

\vspace{1cm}
\end{center}

\begin{abstract}

We revisit the hybrid inflation model gauged by $U(1)_{B-L}$ extension of the Minimal Supersymmetric Standard Model (MSSM) in a no-scale background. Considering a single predictive framework, we study inflation, leptogenesis, gavitino cosmology, and the stochastic background of gravitational waves produced by metastable cosmic strings. The spontaneous breaking of $U(1)_{B-L}$ at the end of inflation produces a network of metastable cosmic strings while, the interaction between $U(1)_{B-L}$ Higgs field and the neutrinos generate heavy Majorana masses for the right-handed neutrinos. The heavy Majorana masses explain the tiny neutrino masses via the seesaw mechanism, a realistic scenario for reheating and non-thermal leptogenesis. We show that a successful non-thermal leptogenesis and a stable gravitino as a dark matter candidate can be achieved for a wide range of reheating temperatures and $U(1)_{B-L}$ symmetry breaking scales. The possibility of realizing metastable cosmic strings in a grand unified theory (GUT) setup is briefly discussed.  We find that a successful reheating with non-thermal leptogenesis and gravitino dark matter restrict the allowed values of string tension to a narrow range $10^{-9} \lesssim G\mu_{CS} \lesssim 8 \times 10^{-6}$, predicting a stochastic gravitational-wave background that lies within the 1-$\sigma$ bounds of the recent NANOGrav 12.5-yr data, as well as within the sensitivity bounds of future GW experiments.

\end{abstract}

%
\section{\large{\bf Introduction}}%

Spontaneous symmetry breaking in grand unified theories (GUTs) can produce variety of topological or non-topological defects \cite{{Vilenkin:1984ib,Vilenkin:2000jqa,Bhattacharjee:1991zm,Hill:1982iq,Kibble:1976sj}}. These defects generically arise from the breaking of a group, $G$, to its subgroup, $H$, such that a manifold of equivalent vacua, $\mathcal{M}$,  $G/H$, exists. Monopoles form when the manifold $\mathcal{M}$ contains non-contractible two-dimensional spheres \cite{tHooft:1974kcl}, cosmic strings when it contains non-contractible loops and domain walls when $\mathcal{M}$ is disconnected \cite{Hindmarsh:1994re}. 
The monopoles can be avoided by inflation which naturally incorporates the GUT scale in supersymmetric hybrid inflation \cite{Dvali:1994ms}. It has been shown for a large class of GUT models that in spontaneous symmetry breaking schemes curing the monopole problem, the formation of cosmic strings cannot be avoided \cite{Jeannerot:2003qv}.

Cosmic strings are interesting messenger from the early universe due to their characteristic signatures in the stochastic gravitational wave background (SGWB). The evidence for a stochastic process at nanohertz frequencies as reported by recent NANOGrav 12.5 year data has been interpreted as SGWB in a large number of recent papers \cite{King:2020hyd,JohnEllis,Buchmuller:2020lbh,King:2021gmj,Ahmed:2021ucx,Vagnozzi:2020gtf,Benetti:2021uea,Lazarides:2021uxv,Samanta:2020cdk,Blasi:2020mfx,Ashoorioon:2022raz}. Relic gravitational waves (GWs) provide a fascinating window to explore the very early universe cosmology \cite{Ahriche:2018rao}.  Cosmic string produce powerful bursts of gravitational radiation that could be detected by interferometric gravitational wave detectors such as LIGO, Virgo, and LISA \cite{LIGOScientific:2019vic, LISA:2017pwj}. In addition, the stochastic gravitational wave background (SGWB) can be detected or constrained by various observations including Big Bang Nucleosynthesis (BBN), pulsar timing experiments and interferometric gravitational wave detectors \cite{Goncharov:2021oub}.

Among the various proposed extensions of the minimal supersymmetric standard model (MSSM), the $U(1)_{B-L}$ is the simplest \cite{Ahmed:2021dvo,Buchmuller:2012wn,Ahmed:2020lua}. Here $B$ and $L$ denote the baryon and lepton numbers, respectively and $B-L$ is the difference between baryon and lepton numbers. As a local symmetry, the $B-L$ group resides in the grand unified (GUT) gauge group $SO(10)$. The spontaneous breaking of $U(1)_{B-L}$  at the end of inflation requires an extended scalar sector, which automatically yields hybrid inflation explaining the inhomogeneities of the CMB. In the $B-L$ breaking phase transition, most of the vacuum energy density is rapidly transferred to non-relativistic $B-L$ Higgs bosons, a sizable fraction also into cosmic strings. The decay of heavy Higgs boson and heavy neutrino leads to an elegant explanation of the small neutrino masses via the seesaw mechanism, explaining the baryon asymmetry via thermal and nonthermal leptogenesis \cite{Fukugita:1986hr,Flanz:1994yx,Vagnozzi:2017ovm,Vagnozzi:2018jhn}. The temperature evolution during reheating is controlled by the interplay between the $B-L$ Higgs and the neutrino sector, while the dark matter originates from thermally produced gravitinos. The embedding of $U(1)_{B-L}$ into a simply-connected group such as $SO(10)$ or Pati-salam symmetry ($SU(4)_C \times SU(2)_L \times SU(2)_R$), produces metastable cosmic strings due to the spontaneous pair creation of a monopole and an anti-monopole. Once the string is cut, the monopoles at the two ends are quickly pulled together due to string tension, forcing them to annihilate. If the string network is sufficiently long-lived, it can generate a stochastic gravitational wave background (SGWB) in the range of ongoing and future gravitational wave (GW) experiments \cite{Buchmuller:2019gfy,Masoud:2021prr}.
 
 Hybrid inflation, in particular, is one of the most promising models of inflation, and can be naturally realized within the context of  supergravity theories. This scenario  is based on the inclusion of two scalar fields \cite{Linde:1993cn}, with the first one realizing the slow-roll inflation and the second one, dubbed the ``waterfall'' field, triggering the end of inflationary epoch. While in the standard hybrid inflationary scenario \cite{Dvali:1994ms,Copeland:1994vg,Linde:1997sj}, the GUT gauge symmetry is broken at the end of inflation, in shifted \cite{Jeannerot:2000sv} and smooth variants \cite{Lazarides:1995vr,Ahmed:2022vlc}, the gauge symmetry breaking occurs during inflation and thus, the disastrous magnetic monopoles and other dangerous topological defects are inflated away.

In this paper we study standard hybrid inflation in the context of supergravity where a no-scale K\"ahler potential is assumed . We consider the framework of MSSM gauge symmetry augmented by a $U(1)_{B-L}$ factor and investigate the implementation of hybrid inflation and its interplay with the issues of non-thermal leptogenesis, gravitino dark matter and stochastic gravitational wave background (SGWB) generated by metastable cosmic string network. For $\mu$ hybrid inflation see Ref \cite{Afzal:2022vjx}. We consider the value of monopole-string-tension ratio from $\sqrt{\lambda} \simeq 7.4$ (metastable cosmic strings) to $\sqrt{\lambda} \simeq 9.0$ (quasi-stable cosmic strings). The parametric space consistent with successful reheating with non-thermal leptogenesis and gravitino dark matter restrict the allowed values of string tension to the range $10^{-9} \lesssim G\mu_{CS} \lesssim 8 \times 10^{-6}$ and predicts a stochastic gravitational wave background (SGWB) that lies within the 1- and 2-$\sigma$ bounds of recent NANOGrav 12.5 years data, as well as the sensitivity bounds of future gravitational wave (GW) experiments. 

The layout of the paper is as follows. In Sec.~2  we describe the basic features of the model including the superfields, their charge assignments, and the superpotential constrained by a
$U(1)_R$ symmetry. The inflationary setup is described  in Sec.~3.  
The numerical analysis is presented in Sec.~4 including the prospects of observing primordial gravity waves, non-thermal leptogenesis, gravitino cosmology and stochastic gravitational wave background (SGWB) generated by metastable cosmic string network. Our conclusion is summarized in Sec~5.
%
\section{\large{\bf Model Description}}%

In this section, we present basic features regarding  the gauge symmetry and the spectrum of the effective model in which the inflationary scenario will be implemented.  The gauge group $U(1)_{B-L}$ is embedded in a grand unified (GUT) gauge group $SO(10)$ and is based on the gauge symmetry
\begin{equation}\label{eq:GBL}
	G_{B-L}=SU(3)_{C}\times{SU(2)_L}\times U(1)_{Y}\times{U(1)_{B-L}}~\cdot
\end{equation}

\noindent  
In addition to the MSSM matter and Higgs superfields, the model supplements six superfields namely; a gauge singlet $S$ whose scalar component acts as inflaton, three right-handed neutrinos $N_{i}^{c}$ and a pair of Higgs singlets $H$ and $\ov{H}$, which are responsible for breaking the gauge group $U(1)_{B-L}$. The charge assignment of these superfields under the gauge symmetry $SU(3)_{C}\times{SU(2)_L}\times U(1)_{Y}\times{U(1)_{B-L}}$ as well as the global symmetries $U(1)_{R}$, $U(1)_B$ and $U(1)_L$ are listed in Table \ref{tab:themodel}. 

The $U(1)_{B-L}$ symmetry is spontaneously broken when the $H$, $\ov{H}$ singlet Higgs super fields acquire vacuum expectation values (VEVs), providing Majorana masses to the right handed neutrinos. The superpotential of the model, invariant under the symmetries listed in Table \ref{tab:themodel}, is given as
\begin{equation}\label{wscalar1}
\begin{split}
W & = \mu H_{u}H_{d}+ y_u {H_{u}}{Q}{u}^c + y_d {H_{d}} {Q}{d}^c +y_{e}{H_{d}}{L}{e}^c + y_{\nu}{H_{u}}{L}{N}^c\\
&+ \kappa S \left(\ov{H}H-M^{2}\right)+ \beta_{ij}^{\prime}\frac{ H H N^{c}N^{c}}{\Lambda}.\;  
\end{split}
\end{equation}
\begin{table}[!t]
	\begin{center}
		\begin{tabular}{c|c|c|c|c}\hline\hline
			{ \sc Superfields}&{\sc Representations}&\multicolumn{3}{c}{\sc
				Global Symmetries}\\\cline{3-5}	%
			&{ \sc under $G_{B-L}$ } & {\hspace*{0.3cm} $U(1)_R$ \hspace*{0.3cm} }
			& {\hspace*{0.3cm}$B$\hspace*{0.3cm}} &{$ L$} \\\hline
			\multicolumn{5}{c}{\sc Matter Fields}\\\hline
			{$e^c_i$} &{$({\bf 1, 1}, 1, 1)$}& $1$&$0$ & $-1$ \\
			{$N^c_i$} &{$({\bf 1, 1}, 0, 1)$}& $1$ &$0$ & $-1$
			\\
			{$L_i$} & {$({\bf 1, 2}, -1/2, -1)$} &$0$&{$0$}&{$1$}
			\\
			{$u^c_i$} &{$({\bf 3, 1}, -2/3, -1/3)$}& $1/2$  &$-1/3$& $0$
			\\
			{$d^c_i$} &{$({\bf 3, 1}, 1/3, -1/3)$}& $1/2$ &$-1/3$& $0$
			\\
			{$Q_i$} & {$({\bf \bar 3, 2}, 1/6 ,1/3)$} &$1/2$ &$1/3$&{$0$}
			\\ \hline
			\multicolumn{5}{c}{\sc Higgs Fields}\\\hline
			
			{$ H_{d} $}&$({\bf 1, 2}, -1/2, 0)$& {$1$}&{$0$}&{$0$}\\
			
			{$ H_{u} $} &{$({\bf 1, 2}, 1/2, 0)$}& {$1$} & {$0$}&{$0$}\\
			\hline
			{$S$} & {$({\bf 1, 1}, 0, 0)$}&$2$ &$0$&$0$  \\
			{$\ov{H}$} &{$({\bf 1, 1}, 0, 1)$}&{$0$} & {$0$}&{$-1$}\\
			{$H$}&$({\bf 1, 1}, 0,-1)$&{$0$}&{$0$}&{$1$}\\
			\hline\hline
		\end{tabular}
	\end{center}
	\caption[]{Superfield contents of the model, the corresponding representations under the local gauge symmetry $G_{B-L}$ and the properties with respect to the extra global symmetries, $U(1)_R$, $U(1)_B$ and $U(1)_L$.}\label{tab:themodel}
\end{table}
\noindent The first line in the above superpotential contains the usual MSSM $\mu$-term and Yukawa couplings supplemented by an additional Yukawa coupling among $L_i$ and $N_i^c$. These Yukawa couplings generate Dirac masses for up and down quarks, charged leptons and neutrinos. The family indices for Yukawa couplings are generally suppressed for simplicity. The first term in the second line is relevent for standard supersymmetric hybrid inflation with $M$ being a GUT scale mass parameter and $\kappa$, a dimensionless coupling constant. The  non-renormalizable term in the second line generates Majorana masses for right handed neutrinos $N_i^c$ and induces the decay of inflaton to $N_i^c$. By virtue of the extra global symmetries, the model is protected from dangerous proton decay operators and $R$-parity violating terms. 

\section{Inflation Potential}
We will compute the effective scalar potential contributions from the $F$- and $D$-sector, radiative corrections as well as the soft supersymmetry breaking terms. The superpotential terms relevant for inflation are 
\begin{equation}\label{wscalar}
W\supset  \kappa S \left(\ov{H}H-M^{2}\right) .
\end{equation}

\noindent  We consider a no-scale structure K\"ahler potential which, after including contributions from the relevant fields in the model, takes the following form

\begin{equation}\label{kahler1}
\begin{aligned}
K =-3 m^{2}_{P} \log \Bigg[T + T^{\ast}- \frac{1}{3 m^{2}_{P}}\left(H H^{\ast} + \bar{H} \bar{H}^{\ast} +S^{\dagger}S\right) &+ \frac{\xi}{3 m^{2}_{P}}\left(H \bar{H} + H^{\ast} \bar{H}^{\ast}\right) \\ &+ \frac{\gamma}{3m^{4}_{P}}\left(S^{\dagger}S\right)^2+....\Bigg],
\end{aligned}
\end{equation}
where $T$, $T^{*}$ are K\"ahler complex moduli fields, $T=\left(u+i\, v\right)$, hence $T+ T^{\ast}=2u$ and $\xi$ is a dimensionless parameter. Here we choose $u=1/2$. For later convenience, we define
\begin{equation}
	\begin{aligned}
	\Delta = \Bigg[T + T^{\ast}- \frac{1}{3 m^{2}_{P}}\left(H H^{\ast} + \bar{H} \bar{H}^{\ast} +S^{\dagger}S\right) &+ \frac{\xi}{3 m^{2}_{P}}\left(H \bar{H} + H^{\ast} \bar{H}^{\ast}\right) \\ &+ \frac{\gamma}{3m^{4}_{P}}\left(S^{\dagger}S\right)^2+....\Bigg]~,
	\end{aligned}
\end{equation}
so that Eq \ref{kahler1} can be written as 
\begin{equation}
	K =-3 m^{2}_{P}\log \Delta.
\end{equation}
The fields carrying $SU(3)_{C}\times{SU(2)_L}\times U(1)_{Y}\times{U(1)_{B-L}}$ quantum numbers are given in Table \ref{tab:themodel} and denoted collectively here with $\phi_i$. The $D$-term potential is given as,
\begin{eqnarray}
V_{D}=\frac{1}{2} D^p_{a} D^p_{a}~,
\end{eqnarray}
where $D_{a}^p$ is defined for $SU(N)$ groups as
\begin{equation*}
	D_{a}^p=-g_{a}K_{,\phi_{i}}\left[t_{a}^p\right]_{i}^{j}\phi_{j}
\end{equation*} 
and for $U(1)$ groups as, 
\begin{equation*}
	D_{a}^p=-g_{a}K_{,\phi_{i}}\left[t_{a}^p\right]_{i}^{j}\phi_{j}-g_{a}q_{i} \varsigma~.
\end{equation*}
Here $K_{,\phi_{i}}\equiv dK/d\phi_{i}$, $\varsigma$ is the Fayet-Iliopoulos coupling constant and $q_i$ are the charges under $U(1)$ group. The $t_{a}^{p}$ are the generators of the corresponding group $G$ and $p = 1, . . . , {\rm dim}(G) $. The $D$-term potential can be written as,
\begin{eqnarray}
V_{D} = \frac{g_{B-L}^{2}}{2\Delta^2}\left[2\mid \bar{H}\mid^{2}-2\mid H\mid^{2}-\xi\left(2H\bar{H}-2\bar{H}H\right)+\left(q_{H}+q_{\bar{H}}\right)\varsigma\right]^2.
\end{eqnarray}
  The $D$-flat potential can be achieve by parametrization of the fields ${H}$ and $\ov{{H}}$. We can rewrite the complex fields in terms of real scalar fields as
\begin{equation} \label{d-flat}
\begin{split}
{H} & = \frac{Y}{\sqrt{2}}e^{\iota\theta}\cos\vartheta, \quad \ov{{H}}=\frac{Y}{\sqrt{2}}e^{\iota\ov{\theta}}\sin\vartheta,
\end{split}
\end{equation}
where the phases $\theta$, $\ov{\theta}$ and $\vartheta$ can be stabilized at 
\begin{eqnarray}
\vartheta=\frac{\pi}{4}\quad \text{and} \quad \theta=\ov{\theta}=0,
\end{eqnarray}
along the $D$-flat direction ($\lvert H \rvert = \lvert \ov{{H}} \rvert =\frac{Y}{2}$). The $F$-term SUGRA scalar potential is given by
\begin{equation}
V_{F}=e^{K/m_{P}^2} \left[\left(K_{i\bar{j}} \right)^{-1}\left(D_{z_{i}}W\right)\left(D_{z_{j}}W\right)^{*}-\frac{3\arrowvert W\arrowvert^{2}}{m_{P}^{2}}\right], \label{Einstein frame SUGRA potential}
\end{equation}
with $z_{i}$ being the bosonic components of the superfields $z%
_{i}\in \{S, H, \bar{H},\cdots\}$, and we have defined
\begin{equation}
D_{z_{i}}W\equiv \frac{\partial W}{\partial z_{i}}+\frac{\partial K}{\partial z_{i}}\frac{W}{m_{p}^{2}}, \ \ \ \ K_{i\bar{j}}\equiv \frac{\partial^{2}K}{\partial z_{i}\partial z_{j}^{*}},
\end{equation}
and $D_{z_{i}^{*}}W^{*}=\left( D_{z_{i}}W\right)^{*}.$
The $F$-term scalar potential during inflation becomes
\begin{eqnarray}
V_{F}(Y,\,|S|)  &=& \frac{\kappa^2}{16} \left( Y^2 - 4 M^2 \right)^2 + \kappa^2 Y^2 |S|^2  
- \kappa^2 M^4 \left(\frac{2}{3} - 4 \gamma \right) \left(\frac{|S|}{m_{p}}\right)^2  \notag \\
&& + \kappa^2 M^4 \left( -\frac{5}{9} + \frac{14 \gamma }{3} + 16 \gamma ^2\right) \left(\frac{|S|}{m_{p}}\right)^4 \cdots.
\end{eqnarray}
Using the $F$-flatness condition, $D_{z_{i}}W = 0$, the minima of potential lies at $Y=2M$ and $S=0$. Along the inflationary trajectory, $Y=0$, the gauge group $U(1)_{B-L}$ is unbroken. After the end of inflation, the spontaneous breaking of the gauge group $U(1)_{B-L}$ yields cosmic strings. Defining dimensionless variable $x\equiv |S|/M$, we obtain the following form of potential along the inflationary trajectory,
\begin{eqnarray}
V_{F}(x) &\simeq& \kappa^2 M^4 \left( 1 -\left(\frac{2}{3} - 4 \gamma \right)\left(\frac{Mx}{m_{p}}\right)^2 + \left( -\frac{5}{9} + \frac{14 \gamma }{3} + 16 \gamma ^2\right)\left(\frac{M x }{m_{p}}\right)^4 +  \cdots\right).  \label{SHIpot} 
\end{eqnarray}
 The action of our model for non-canonically normalized field $x$ is given by
  \begin{equation}
	\begin{split}
		\mathcal{A}= \int dx^{4}\sqrt{-g}\left[\frac{m^2_{p}}{2}\mathcal{R}- K^{i}_{j}\partial_{\mu} x^{i}\partial^{\mu} x^{j} -V({x})\right].
	\end{split}
\end{equation}
Introducing a canonically normalized field $\chi$ that satisfies
\begin{equation}\label{jphi}
\begin{split}
\left(\frac{d\chi}{dx}\right)^{2} =\frac{3 \frac{\gamma  M^2}{m_{p}^2} x^2 \left(\frac{M^2}{m_{p}^2} x^2-12\right)+9}{\left(\frac{M^2}{m_{p}^2} x^2 \left(  \frac{\gamma M^2}{m_{p}^2} x^2-1\right)+3\right)^2}\sim1.
\end{split}
\end{equation}
Since $\gamma M^2\ll m^2_{p}$, integrating Eq. \eqref{jphi} in this limit, we obtain the canonically normalized field $\chi$ as a function of $x$. The canonically normalized potential as a function of $\chi$ can be written as,
\begin{eqnarray}
V_F (\chi)
&\simeq&
\kappa ^{2}M^{4}\left( 1  -\left(\frac{2}{3} - 4 \gamma \right) \left(\frac{M}{m_{p}}\right)^2 \chi^{2}+ \left( -\frac{5}{9} + \frac{14 \gamma }{3} + 16 \gamma ^2\right) \left(\frac{M}{m_{p}}\right)^4 \chi^4+\cdots\right).
\label{scalarpot}
\end{eqnarray}
The effective scalar potential including the well-known radiative corrections and soft SUSY breaking terms, can be expressed as
\begin{eqnarray}
V(\chi) &\simeq&V_{F}+V_{D}+V_{CW}+V_{Soft} \\
&\simeq&\kappa^2 M^4 \left(  1  -\left(\frac{2}{3} - 4 \gamma \right) \left(\frac{M}{m_{p}}\right)^2 \chi^{2}+ \left( -\frac{5}{9} + \frac{14 \gamma }{3} + 16 \gamma ^2\right) \left(\frac{M}{m_{p}}\right)^4+\right.  \notag  \\
&& + \left.\frac{\kappa ^2}{8\pi ^2}F(\chi) + a \left(\frac{m_{3/2} \chi }{\kappa M}\right)+\left(\frac{M_{S}\chi}{\kappa M}\right)^2 \right),  \label{SHIpot} 
\end{eqnarray}
with
\begin{equation}
	a = 2|A-2| \cos \left[\arg S + \arg (1-A)\right],
\end{equation}  
and
\begin{equation}
F(\chi)=\frac{1}{4}\left(\left(\chi^4+1\right)\log\left(\frac{\chi^4-1}{\chi^4}\right) + 2 \chi^2\log\left(\frac{\chi^2+1}{\chi^2 - 1}\right) + 2 \log\left(\frac{\kappa^2 M^2 \chi^2}{Q^2}\right) - 3 \right).
\end{equation}
Here, $Q$ is the renormalization scale, $a$ and $M_S$ are the coefficients of soft SUSY breaking linear and mass terms for $S$, respectively, and $m_{3/2}$ is the gravitino mass. For simplicity, we set $M_{S}=m_{3/2}$ and assume a suitable initial condition for $\arg S$ to be stabilized at zero and take $a$ to be constant during inflation (for details see Ref \cite{Buchmuller:2014epa}).

\section{Analysis}

In this section, we analyze the implications of the model and discuss its predictions regarding the various cosmological observables. We pay particular attention to inflationary predictions and stochastic gravitational waves (GW) spectrum consistent with leptogenesis and gravitino cosmology.

\subsection{Inflationary Predictions}
 The inflationary slow-roll parameters can be expressed in terms of $\chi$ as
\begin{equation}
	\epsilon=\dfrac{1}{4}\left(\frac{m_{p}}{M}\right)^2\left(\frac{V^{\prime}\left(\chi\right)}{V(\chi)}\right)^{2}, \quad\quad \eta=\dfrac{1}{2}\left(\frac{m_{p}}{M}\right)^2\left(\frac{V^{\prime\prime}\left(\chi\right)}{V(\chi)}\right),
\end{equation}
\begin{equation}
	s^{2}=\dfrac{1}{4}\left(\frac{m_{p}}{M}\right)^4\left(\frac{V^{\prime}\left(\chi\right)V^{\prime\prime}\left(\chi\right)}{V(\chi)}\right),
\end{equation}
\noindent where a prime denotes a derivative with respect to $\chi$. The tensor-to-scalar ratio $r$, the scalar spectral index $n_{s}$, and the running of the spectral index $\frac{dn_{s}}{d\ln k}$ are given by
\begin{equation}
	r\simeq16 \epsilon \quad{,}\quad n_{s}\simeq 1+2\eta-6\epsilon \quad{,}\quad \frac{dn_{s}}{d\ln k}\simeq 16\epsilon\eta-24\epsilon^{2}+2s^{2}.
\end{equation}
The number of e-folds is given by \cite{Garcia-Bellido:1996egv},
\begin{equation}
	N_{l}=2\left(\frac{M}{m_{p}}\right)^2\int_{\chi_{e}}^{\chi_{l}}\left(\frac{V\left(\chi\right)}{V^{\prime}(\chi)}\right) d\chi= 54+\frac{1}{3}\ln\left[\frac{T_{r}}{10^{9} {\rm GeV}}\right]+\frac{1}{3}\ln\left[\frac{V(\chi_{l})^{1/4}}{10^{16} {\rm GeV}}\right]~,
\end{equation}
where $l$ denotes the comoving scale after crossing the horizon, $\chi_{l}$ is the field value at $l$, $\chi_e$ is the field value at the end of inflation, (i.e., when  $\epsilon=1$), and  $T_{r}$ is the reheating temperature which will be discussed in the following section. The amplitude of curvature perturbation $\Delta_{R}$ is given by
\begin{equation}
	\Delta_{R}^{2}=\frac{V\left(\chi\right)}{24 \pi^{2} \epsilon\left(\chi\right)}.
\end{equation}
\begin{figure}[t]
	\centering \includegraphics[width=7.93cm]{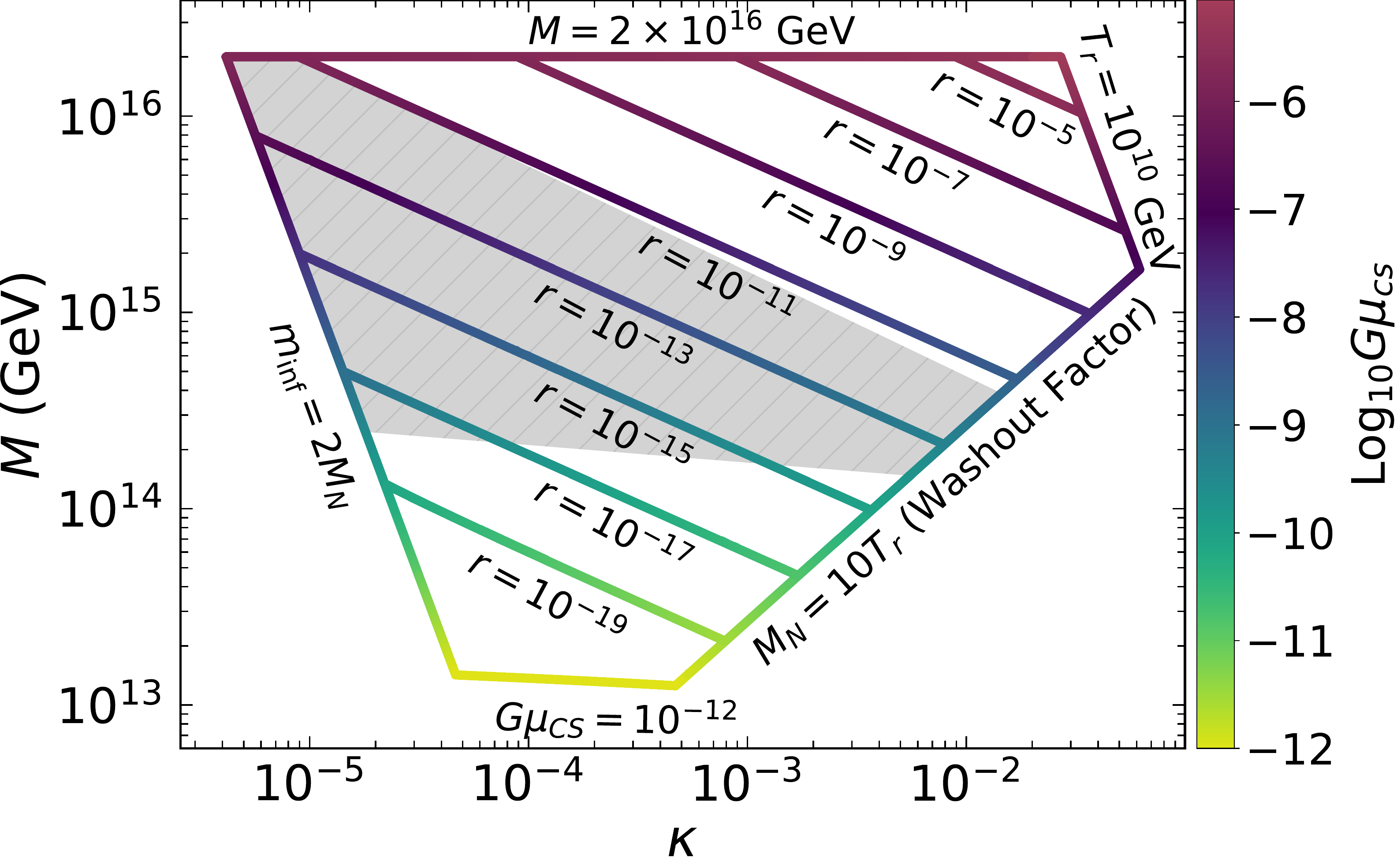} 
	\centering \includegraphics[width=7.93cm]{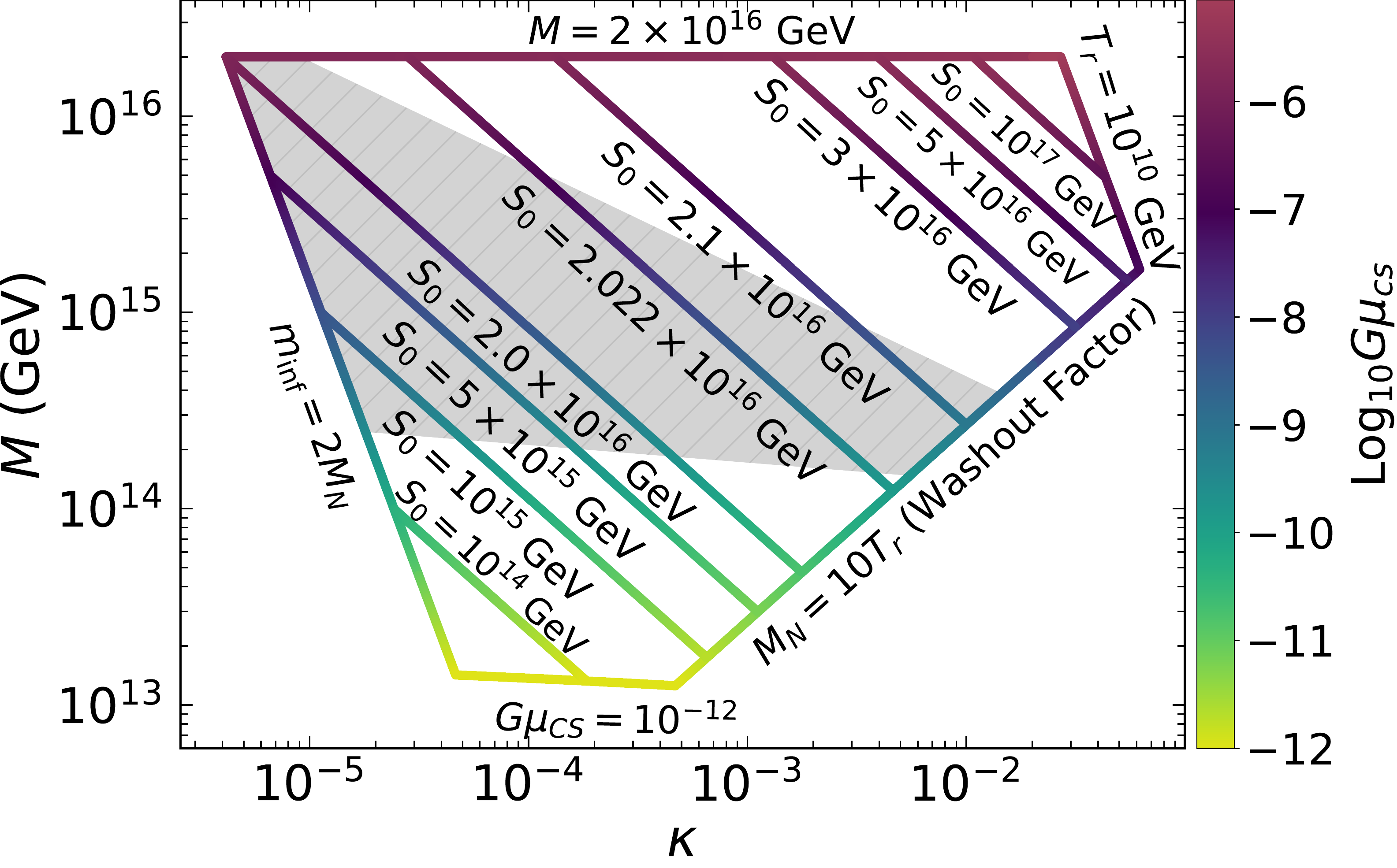}
	\caption{Contours of tensor to scalar ratio $r$ (left panel) and the field $S_0$ (right panel) in the $\kappa-M$ plane, where $M$ is the $B-L$ guage symmetry breaking scale. The boundary is drawn for different constraints shown. The color bar on the right displays the range of string tension parameter $G \mu_{CS}$. The shaded region represents the parametric space that is consistent with gravitino dark matter.}
	\label{fig1}
\end{figure}
The results of our numerical calculations are presented in Fig. \ref{fig1} where the variation of parameters is shown in the $\kappa-M$ plane. In our analysis, the scalar spectral index is fixed at the central value of Planck's bounds $n_{s}=0.9655$. To keep the SUGRA expansion, parameterized by $\gamma$, under control we impose $S_{0}\leq m_{p}$. We restrict $M\leq 2\times10^{16}$ GeV and $T_{r}\leq 10^{10}$ GeV to avoid the gravitino problem. We further restrict our numerical results by imposing the following conditions
\begin{equation}
	m_{inf}=2M_{N}, \qquad M_{N}=10 T_{r},
\end{equation}
which ensure successful reheating with non-thermal leptogenesis. The boundary curves in Fig. \ref{fig1} represent; $M = 2 \times 10^{16}$ GeV, $T_r = 10^{10}$ GeV, $m_{\text{inf}} = 2 M_N$, $M_N = 10 T_r$ and $G \mu = 10^{-12}$ constraints.  The left panel in Fig. \ref{fig1} shows the variation of tensor to scalar ratio $r$, whereas, the right panel shows the variation of the field value $S_{0}$. The color bar depicts the range of string tension $G\mu_{CS}$ obtained in our model. It should be noted that the parameter $\gamma$ which controls the SUGRA corrections, makes this model more predictive than the standard hybrid model of inflation. Using leading order slow-roll approximation, we obtain the following analytical expressions for $n_{s}$ in the small $\kappa$ limit,
\begin{equation}
	n_s = 1 - 2 \left( \frac{2}{3} - 4 \gamma \right).
\end{equation}
It can readily be check that for $\gamma = 0.162292$, we obtain $n_{s} \sim 0.9655$ which is in excellent agreement with the numerical results displayed in Fig. \ref{fig1}. The above equation therefore gives a valid approximation of our numerical results. For the scalar spectral index $n_s$ fixed at Planck's central value ($0.9655$), we obtain the following ranges of parameters 
\begin{gather}
	\nonumber
	4.2 \times 10^{-6} \lesssim \kappa \lesssim 6.2 \times 10^{-2}, \\ \nonumber
	(1.3 \times 10^{13} \lesssim M \lesssim 2.0 \times 10^{16}) ~ \text{GeV}, \\ \nonumber
	(2 \times 10^{16} \lesssim S_0 \lesssim 2 \times 10^{17}) ~ \text{GeV}, \\ \nonumber
	7.1 \times 10^{-23} \lesssim r \lesssim 10^{-4}, \\ 
	10^{-12} \lesssim G \mu_{CS} \lesssim 8 \times 10^{-6}. 
\end{gather}
Using the Plank's normalization constraint on $\Delta_{R}$, we obtain the following explicit dependence of $r$ on $\kappa$ and $M$ 
\begin{equation}
	r \simeq \frac{2 \kappa^2}{3 \, \pi^2 \Delta_{R}^2} \left(\frac{M}{m_P}\right)^4,
\end{equation}
which explains the behavior of tensor to scalar ratio $r$ in $\kappa-M$ plane. It can readily be checked that for $\kappa \simeq 4.65 \times 10^{-5}$ and $M \simeq 1.42 \times 10^{13}$ GeV, the above equation gives $r \simeq 7.9 \times 10^{-23}$. On the other hand, $\kappa \simeq 2.7 \times 10^{-2}$ and $M \simeq 2.0 \times 10^{16}$ GeV gives $r \simeq 10^{-4}$. These approximate values are very close to the actual values obtained in the numerical calculations.

\subsection{\large{\bf Reheating with non-thermal leptogenesis }}\label{sec4}

At the end of inflation epoch, the vacuum energy is transfered to the energies of  coherent oscillations of the  inflaton $S$ and the scalar field $\theta=(\delta H+\delta\bar{H})/\sqrt{2}$ whose  decays give rise to the radiation in the universe. The inflaton decay to right handed neutrino is induced by the superpotential term
\begin{equation}
W \supset \beta_{ij}^{\prime}\frac{ H H N^{c}N^{c}}{\Lambda} ,\label{Infnu1}
\end{equation}
where $\beta_{ij}^{\prime}$ is a coupling constant and $\Lambda$ represents a high cut-off scale (in a string model this could be identified with  the compactification scale). Heavy Majorana masses for the right-handed neutrinos are provided by the following  term 
\begin{equation}
M_{\nu^c_{ij}}=\beta_{ij}^{\prime}\frac{\langle H \rangle \langle H \rangle }{\Lambda}~\cdot 
\end{equation}
Also, Dirac neutrino masses of the order of the electroweak scale  are obtained from the  tree-level superpotential term  ${y_{\nu}}_{ij}\,N_{i}^c\,L_{j}\,H_{u}\to {m_{\nu_D}}_{ij}N N^c $   given in~(\ref{wscalar1}).  Thus, the neutrino sector is
\be\label{Infnu2}
W\supset  {m_{\nu_D}}_{ij}N_iN_j^c+ M_{\nu^c_{ij}}N_i^cN_j^c.
\ee  
 The small neutrino masses supported by neutrino oscillation experiments, are obtained by integrating out the heavy right-handed neutrinos and read as
\begin{equation}
{m_{\nu_D}}_{\alpha\beta}=-\sum_{i}{y_{\nu}}_{i\alpha}{y_{\nu}}_{i\beta}\frac{v_{u}^2}{M_i}~\cdot \label{mneu1}
\end{equation}
The neutrino mass matrix ${m_{\nu_D}}_{\alpha\beta}$ can be diagonalized by a unitary matrix $U_{\alpha i}$ as ${m_{\nu_D}}_{\alpha\beta} =  U_{\alpha i} U_{\beta i} m_{\nu_D}$, where $m_{\nu_D}$ is a diagonal
mass matrix $m_{\nu_D} = {\rm diag}(m_{\nu_{1}}, m_{\nu_{2}}, m_{\nu_{3}})$ and $M_{i}$ represent the eigenvalue of mass matrix  $M_{\nu^c_{ij}}$.

 The lepton asymmetry is generated (inducing also baryon asymmetry~\cite{Fukugita:1986hr,Flanz:1994yx})
through right-handed neutrino decays. The lepton number density
to the entropy density  in the limit $T_r < M_{1}\equiv M_{N}\leq m_{\text{inf}} /2 \leq M_{2,3}$ is defined as
\begin{equation}
\frac{n_{L}}{s}\sim \frac{3}{2}\frac{T_{r}}{m_{\text{inf}}}\epsilon_{cp}~,
\end{equation}
where $\epsilon_{cp}$ is the CP asymmetry factor and is generated from the out of equilibrium decay of lightest right-handed neutrino and is given by \cite{Hamaguchi:2002vc},
\begin{equation}
\epsilon_{cp}=-\frac{3}{8\pi}\frac{1}{\left({y_{\nu}}{y_{\nu}}^{\dagger}\right)_{11}}\sum_{i=2,3}\operatorname{Im} \left[\left({y_{\nu}}{y_{\nu}}^\dagger\right)_{1i}\right]^2\frac{M_{N}}{M_i},
\end{equation}
and $T_{r}$ is reheating temperature which can be as estimated as
\begin{eqnarray}	
T_r \simeq \sqrt[4]{\frac{90}{\pi^2 g_{\star}}} \sqrt{\Gamma \, m_P}~,
\label{reheat}
\end{eqnarray} 
where $g_{\star}$ is $228.75$ for MSSM. The $\Gamma$ is the decay width for the inflaton decay into right-handed neutrinos and is given by \cite{Hamaguchi:2002vc}
\begin{equation}
\Gamma \left({\rm inf} \rightarrow N_{i}^c N_{j}^c \right) = \frac{1}{8 \pi}\left(\frac{M_{N}}{M}\right)^2 \, m_{\text{inf}}  \left(  1 - \frac{4 M_{N}^2}{m_{\text{inf}}^2} \right)^{1/2},
\end{equation}
with the inflaton mass given by
\begin{equation}
m_{\text{inf}} = \sqrt{2\kappa^2M^2+M_{S}^2}.
\end{equation}
Assuming a normal hierarchical pattern of light neutrino masses, the CP asymmetry factor, $\epsilon_{cp}$, becomes 
\begin{equation}
\epsilon_{cp} = \frac{3}{8\pi}\frac{M_{N} m_{\nu_{3}}}{v_{u}^2}\delta_{\rm eff}, 
\end{equation}
where $m_{\nu_3}$ is the mass of the heaviest light neutrino, $v_{u}=\langle H_u \rangle $ is the VEV of the up-type electroweak Higgs and $\delta_{\rm eff}$ is the CP-violating phase. The experimental value of lepton asymmetry is estimated as \cite{Planck:2018vyg},
\begin{eqnarray}
\mid n_L/s\mid\approx\left(2.67-3.02\right)\times 10^{-10}.
\end{eqnarray}
\begin{figure}[!htb]
	\centering \includegraphics[width=7.93cm]{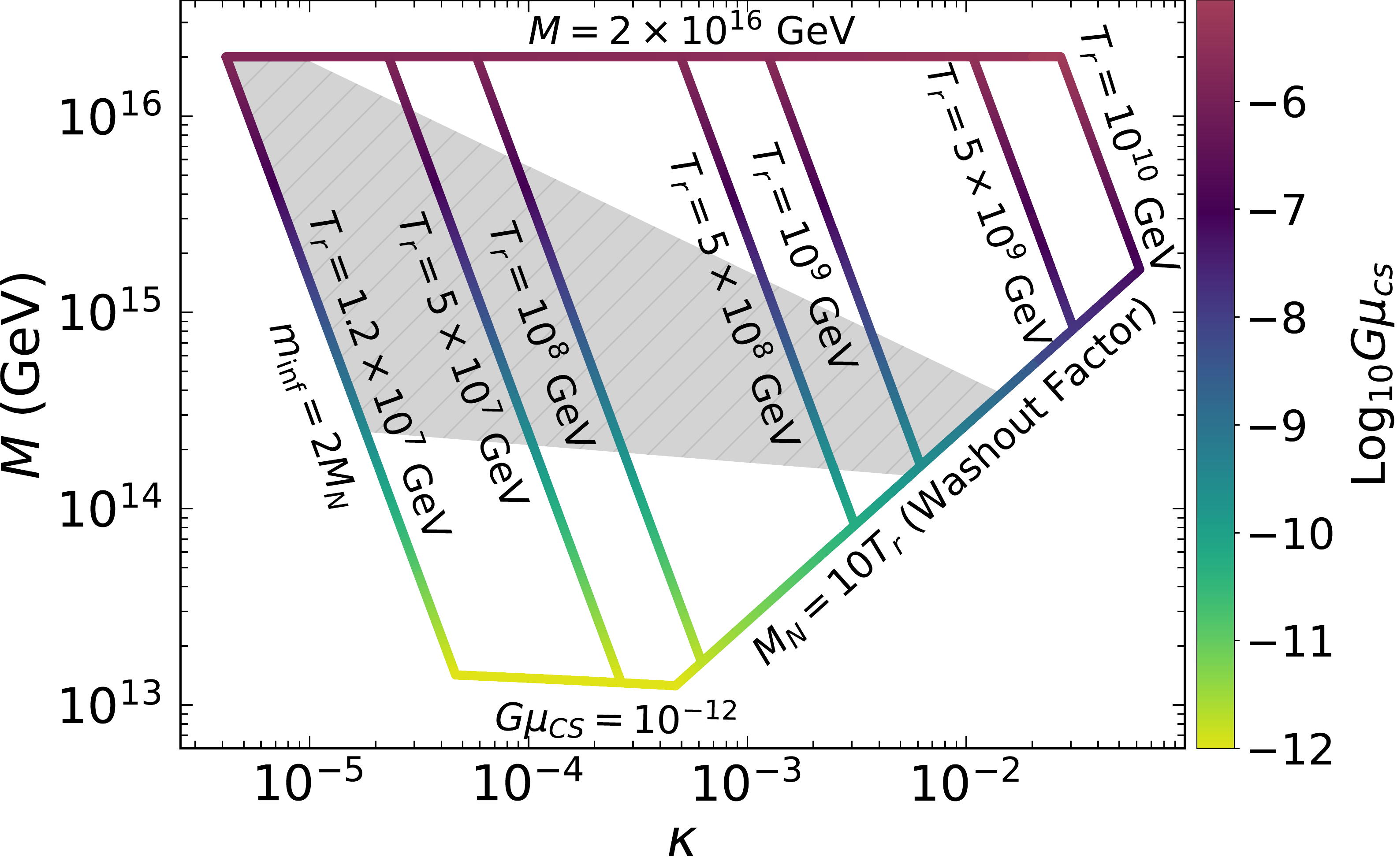} \\ \vspace*{10pt}
	\centering \includegraphics[width=7.93cm]{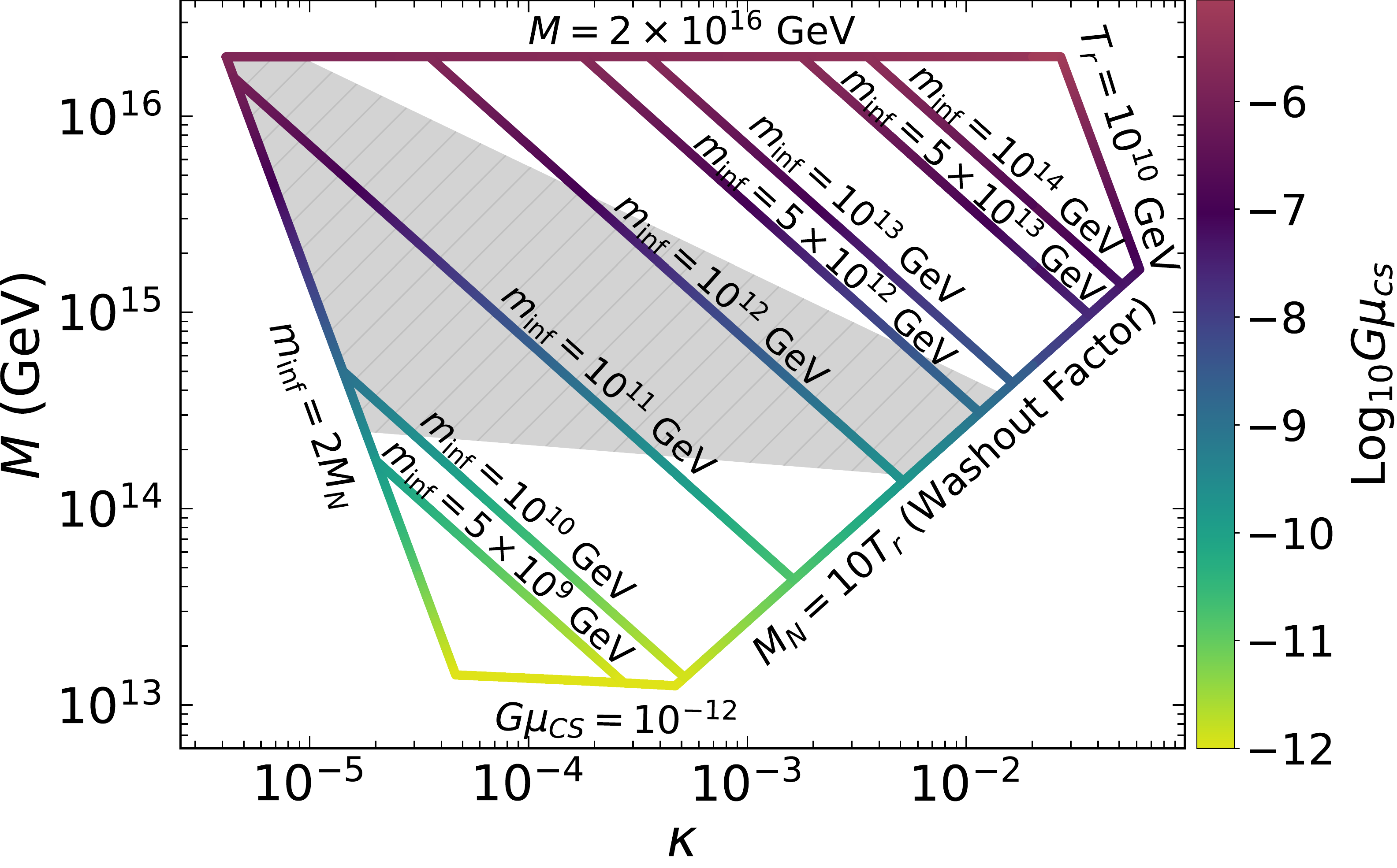}
	\centering \includegraphics[width=7.93cm]{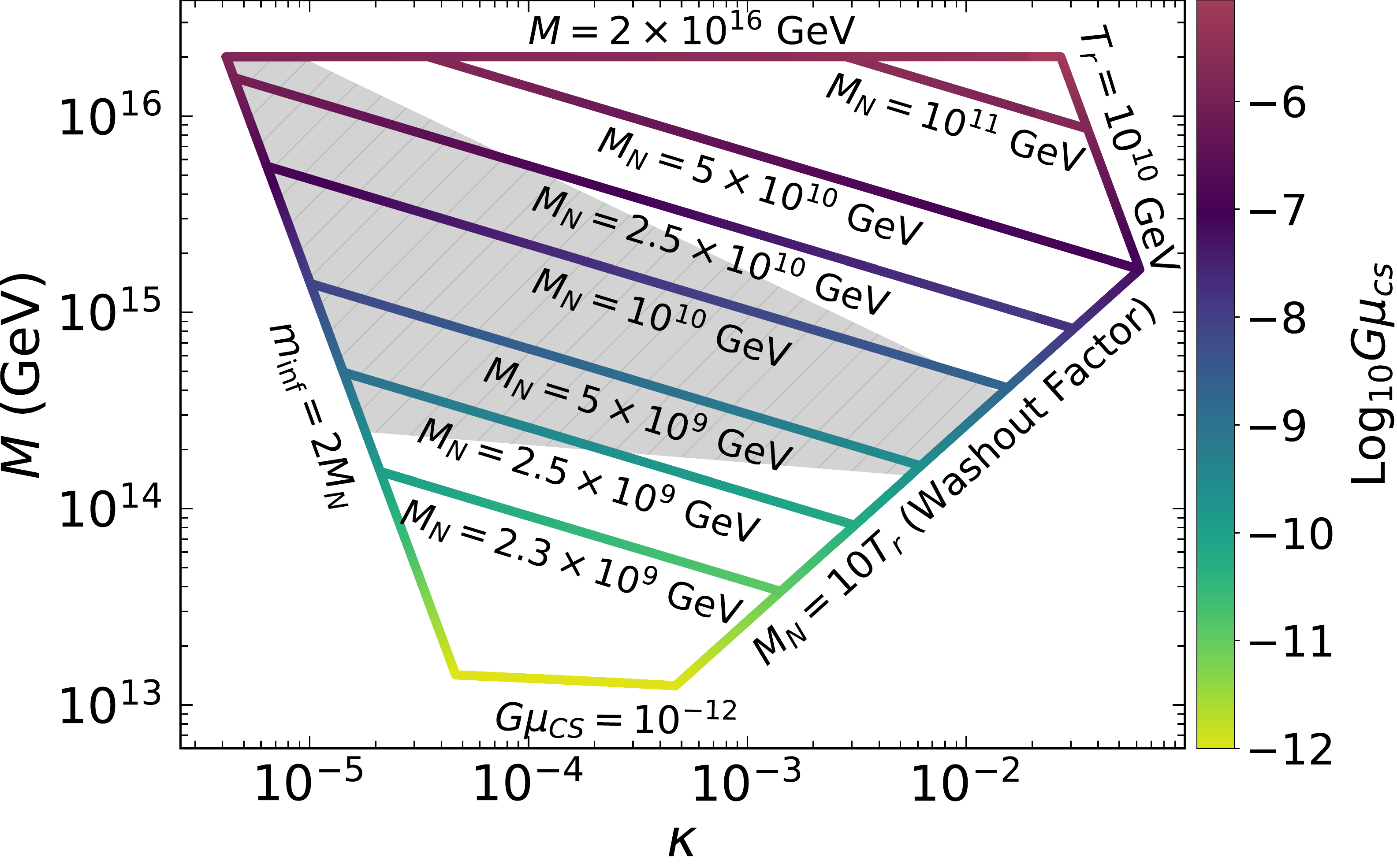}
	\caption{Contours of the reheat temperature $T_r$ (top), the inflaton mass $m_{\text{inf}}$ (bottom left) and the right handed neutrino mass $M_N$ (bottom right), in the $\kappa-M$ plane, where $M$ is the $B-L$ guage symmetry breaking scale. The boundary is drawn for different constraints shown. The color bar on the right displays the range of string tension parameter $G \mu_{CS}$. The shaded region represents the parametric space that is consistent with gravitino dark matter.}
	\label{fig2}
\end{figure}
In the numerical estimates discussed below we take $m_{\nu_3} = 0.05$ eV, $|\delta_{\rm eff}|=1$ and $v_u = 174$ GeV, while assuming large $\tan \beta $.
The non-thermal production of lepton asymmetry, $n_{L}/s$, is given by the following expression  
\begin{equation}
\frac{n_L}{s} \lesssim 3 \times 10^{-10} \frac{T_r}{m_{\text{inf}}}\left(\frac{M_{N}}{10^6 \text{ GeV}}\right)\left(\frac{m_{\nu_3}}{0.05 \text{ eV}}\right) \label{nls},
\end{equation}
with $M_{1} \gg T_r $. Using the experimental value of $n_L/s\approx 2.5\times 10^{-10}$ with Eq. \eqref{reheat} and \eqref{nls}, we obtain the following lower bound on $T_r$,
\begin{equation}
T_r \gtrsim 1.9 \times 10^7 \text{ GeV}  \left(\frac{m_{\text{inf}}}{10^{11}\text{ GeV}}\right)^{3/4} \left(\frac{10^{16} \, \text{GeV}}{M_{N}}\right)^{1/2}\left(\frac{m_{\nu_3}}{0.05 \text{ eV}}\right)^{1/2} \label{lepto}.
\end{equation}
A successful baryogenesis is usually generated through the sphaleron processe where an initial lepton asymmetry, $n_L/s$, is partially converted into a baryon asymmetry  \cite{Khlebnikov:1988sr,Harvey:1990qw}. However, the right handed neutrinos produced in inflaton decays are highly boosted, which affects the estimate of the final baryon asymmetry as given in \cite{Buchmuller:2019gfy}. Eq. \eqref{lepto} is used in our numerical analysis to calculate inflationary predictions which are consistent with leptogenesis and baryogenesis.

Fig. \ref{fig2} shows the contours of reheating temperature $T_r$ (top), inflaton mass $m_{\text{inf}}$ (bottom left) and right handed neutrino mass $M_N$ (bottom right) in $\kappa-M$ plane. We obtain these parameters in the following ranges 
\begin{gather}
	\nonumber
	(10^{7} \lesssim T_r \lesssim 10^{10}) ~ \text{GeV}, \\ \nonumber
	(4.7 \times 10^{8} \lesssim M_N \lesssim 5.3 \times 10^{11}) ~ \text{GeV}, \\ 
	(9.4 \times 10^{8} \lesssim m_{\text{inf}} \lesssim 7.6 \times 10^{14}) ~ \text{GeV}. \\ \nonumber
\end{gather}
The color bar on the right displays the range of string tension parameter $G\mu_{CS}$, while the shaded region corresponds to the stable gravitino, as discussed in the following section. 

\subsection{Gravitino Dark Matter}
An important constraint on the reheat temperature $T_{r}$ arises, when gravitino cosmology is taken into account, that depends on the SUSY breaking mechanism and the gravitino mass $m_{3/2}$. As noted in \cite{Ahmed:2021dvo,Lazarides:2020zof}, one may consider the case of 
\\ $\alpha)$ a stable LSP gravitino; \\ $\beta)$ unstable long-lived gravitino with mass
$m_{3/2} < 25$ TeV;  \\ $\gamma)$ unstable short-lived gravitino with mass $m_{3/2} > 25$ TeV.

We first consider the case of stable gravitino, in which case it is the lightest SUSY particle (LSP) and assuming it is thermally produced, its relic density is estimated to be \cite{Bolz:2000fu}
 \begin{equation}\label{omega}
 \Omega_{3/2} h^{2}=0.08\left(\frac{T_{r}}{10^{10} \; \text{GeV} }\right)\left(\frac{m_{3/2}}{1 \; \text{TeV}}\right)\left(1+\frac{m_{\tilde{g}^{2}}}{3m_{3/2}^{2}}\right)~,
 \end{equation}
 where $m_{\tilde{g}}$ is the gluino mass parameter and $h$ is the present Hubble parameter in units of $100$ km $\text{sec}^{-1} \text{Mpc}^{-1}$  and $\Omega_{3/2}= \rho_{3/2}/\rho_c $. \footnote{$\rho_{3/2}$ and $\rho_c$ are the present energy density of the gravitino and the critical energy density of the present universe, respectively.} \footnote{Eq. (\ref{omega}) contains only the dominant QCD contributions for the gravitino production rate. In principle there are extra contributions descending from the electroweak sector as mentioned in \cite{Pradler:2006qh}, \cite{Rychkov:2007uq} and recently revised in \cite{Eberl:2020fml}. If we consider these type of contributions in our analysis, we estimate that (depending on gaugino universality condition) our results  will deviate $\sim{(10-15)\%}$.}. A stable  LSP  gravitino requires $m_{\tilde{g}}>m_{3/2}$ while current LHC  bounds on the gluino mass are around $2.2$ TeV \cite{Vami:2019slp}.  It is found from Eq. \ref{omega} that the overclosure limit $\Omega_{3/2}<1$ puts a severe upper bound on the reheating temperature $T_{r}$, depending on the gravitino mass $m_{3/2}$. Here, we have omitted the contribution from the decays of squarks and sleptons into gravitinos. 
  \begin{figure}[!ht] 
  	\centering \includegraphics[width=12cm]{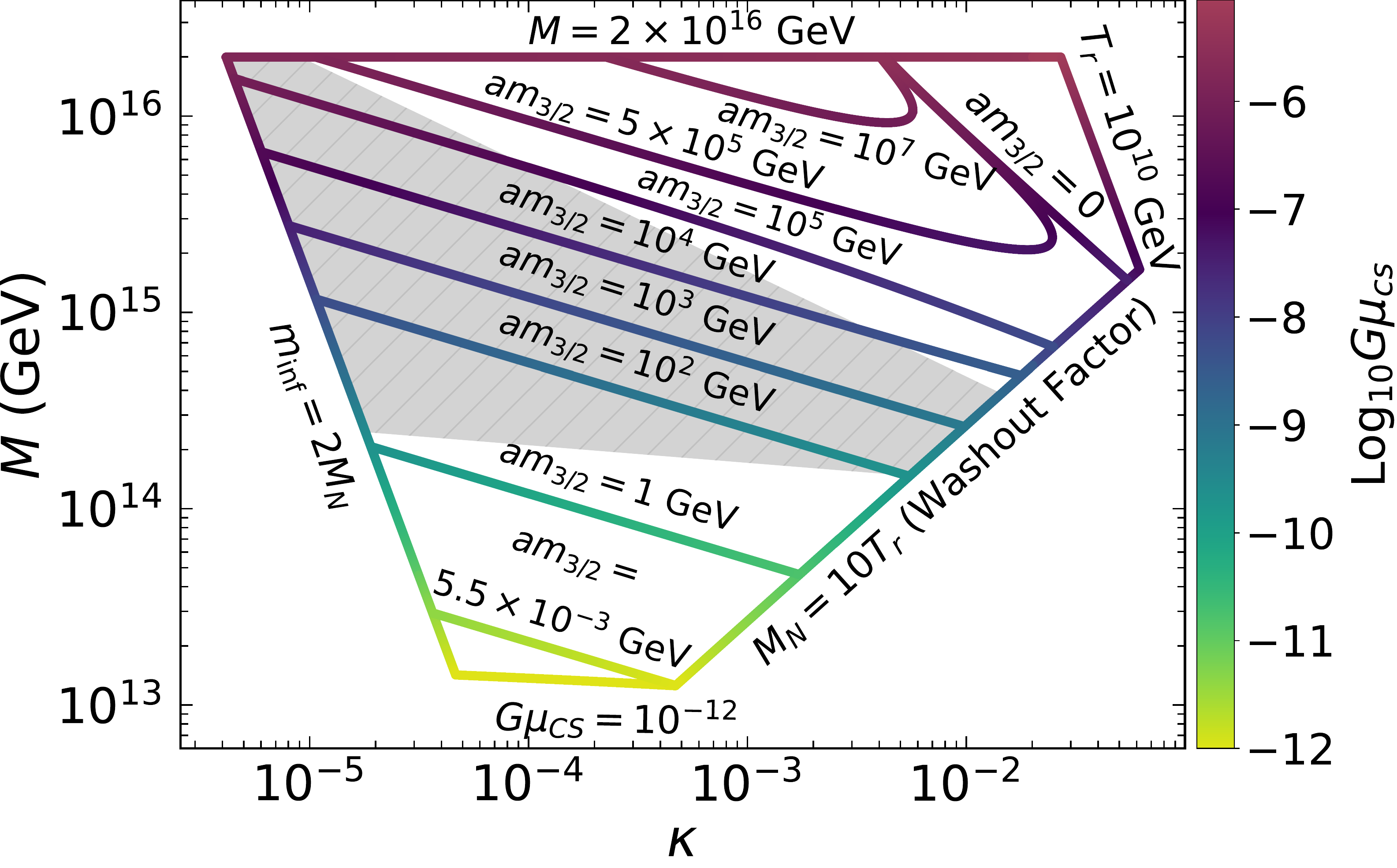}
  	\caption{Contours of the gravitino mass $m_{3/2}$ in the $\kappa-M$ plane, where $M$ is the $B-L$ guage symmetry breaking scale. The boundary is drawn for different constraints shown. The color bar on the right displays the range of string tension parameter $G \mu_{CS}$. The gray shaded region corresponds to the parametric space where the gravitino is LSP $(m_{\tilde{g}}>m_{3/2})$.}
  	\label{grav}
  \end{figure}Using the lower bound of relic abundance $\Omega^{2}_{h}=0.144$ \cite{Akrami:2018odb} in the above equation, we display a gray shaded region in Fig. \ref{grav} that satisfies the condition $(m_{\tilde{g}}>m_{3/2})$ and hence, in this region, gravitino is the lightest supersymmetric particle (LSP) and acts as a viable dark matter candidate. For gravity mediated SUSY breaking, the constraints on gravitino mass and reheat temperature from BBN  are \cite{Kawasaki:2004qu}

 \begin{equation} \label{eqlong}
 \begin{split}
 T_{r}&  \lesssim 10^{7} {\rm GeV \quad for }\quad m_{3/2}= (100-5000)\; {\rm GeV }\; ,\\
 T_{r}& \sim (10^7 - 2.5\times10^{9}) \;{\rm GeV } \quad {\rm for} \quad m_{3/2}\geq 5000\; {\rm GeV. }
 \end{split}
 \end{equation} 
The shaded region in above Fig. \ref{grav} describes the gravitino dark matter and the value of gravitino mass varies in the range $1.5 \; \text{GeV} \lesssim m_{3/2} \lesssim 4.2 \times 10^5 \; \text{GeV}$ with reheat temperature $T_r \gtrsim  10^7 \; \text{GeV}$. However, the gravitino mass in the range $100 \; \text{GeV} \lesssim m_{3/2} \lesssim 5000 \; \text{GeV}$ requires a reheat temperature $T_r < 10^7 \; \text{GeV}$, which is not achieved in our model and therefore, this small range of $m_{3/2}$ is ruled out by BBN.

An unstable gravitino, could be either long-lived or a short-lived. The lifetime of a long-lived gravitino with mass $m_{3/2}<25$ TeV is about  $\tilde{\tau}\gtrsim 1 \; \text{sec}$. A long-lived gravitino leads to the cosmological gravitino problem \cite{Khlopov:1984pf} that originates due to the fast decay of gravitino which may  affect the light nuclei abundances and thereby ruin the success of BBN theory. To avoid this problem, one has to take into account the BBN bounds (Eq. \ref{eqlong}) on the reheating temperature. Nevertheless for all range of $5000\leq m_{3/2}\leq 25000$ GeV, a long-lived gravitino scenario is viable and  consistent with the BBN bounds \eqref{eqlong}. 
 
 For short-lived gravitino, the BBN bounds on the reheating temperature are not effective and gravitino decays into the LSP neutralino $\tilde{\chi}_{1}^{0}$, for which the abundance is given by
 \begin{equation}\label{eqa}
 \Omega_{\tilde{\chi}_{1}^{0}}h^{2}\simeq 2.8\times10^{11}\times Y_{3/2}\left(\frac{m_{\tilde{\chi}_{1}^{0}}}{1 \; \text{TeV}}\right),
 \end{equation}
 where $Y_{3/2}$ is the gravitino Yield and is defined as,
 \begin{equation}\label{eqb}
 Y_{3/2}\simeq2.3\times10^{-12}\left(\frac{T_{r}}{10^{10} \; \text{GeV}}\right).
 \end{equation}
Since the LSP neutralino density produced by gravitino decay should not exceed the observed DM relic density, choosing the upper bound of relic abundance $\Omega_{\tilde{\chi}_{1}^{0}}h^{2} = 0.126$ and using equations (\ref{eqb}) and (\ref{eqa}), we find a relation between the reheating temperature $T_{r}$ and $m_{\tilde{\chi}_{1}^{0}}$,
given by
 \begin{eqnarray}\label{eqc}
 m_{\tilde{\chi}_{1}^{0}}\simeq19.6\left(\frac{10^{11} \; \text{GeV}}{T_{r}}\right)~.
 \end{eqnarray}
 For gravity mediation scenario $ m_{\tilde{\chi}_{1}^{0}}\geq18$ GeV \cite{Hooper:2002nq}, which is easily satisfied in the current model. Therefore, the short-lived gravitino scenario is also a viable possibility  in this  model. The region above the shaded area in Fig. \ref{grav} corresponds to short-lived gravitino. Finally, we obtain the following ranges of string tension $G\mu_{CS}$ and gravitino mass for stable and unstable gravitino consistent with BBN bounds,
\begin{gather} \nonumber
10^{-9} \lesssim G\mu_{CS} \lesssim 8 \times 10^{-6}, \\ 
	(-3.2 \times 10^{9} \lesssim am_{3/2} \lesssim 5 \times 10^{8}) ~ \text{GeV}.
\end{gather}
In next sub section, we analyze stochastic gravitational waves (GW) spectrum, consistent with leptogenesis and gravitino cosmology.
\subsection{Gravitational Waves From Cosmic Strings}
The superposition of GW sources, such as inflation, cosmic strings and phase transition, would generate a stochastic GW background (SGWB). The tensor perturbations upon horizon re-entry give rise to the inflationary SGWB \cite{ Vagnozzi:2020gtf, Benetti:2021uea,Caprini:2015tfa,Kuroyanagi:2020sfw} which imprint a distinctive signature in the CMB $B$-mode polarization. The amplitude and scale dependence of the inflationary SGWB is parameterized via the tensor-to-scalar ratio $r$ and the tensor spectral index $n_T$, which satisfy the inflationary consistency relation $r = -8 n_T$ \cite{Liddle:1993fq}, within single-field and hybrid slow-roll models. Since $r \geq 0$, this requires $n_T \leq 0$ (red spectrum)\cite{BICEP2:2018kqh}. With current constraints on tensor-to-scalar ratio $r$, the amplitude of the inflationary SGWB on PTA and interferometer scales is far too small to be detectable by these probes and would instead require a strong blue tilted ($n_T > 0$) primordial tensor power spectrum  \cite{ Vagnozzi:2020gtf}. 
For a detailed study on SGWBs from first-order phase transition associated with the spontaneous $U(1)_{B-L}$ gauge symmetry breaking, see Refs \cite{Jinno:2016knw,Hasegawa:2019amx,Haba:2019qol,Dong:2021cxn}.

 In this section, we study SGWB spectra produce by the decay of cosmic string network \cite{King:2020hyd,JohnEllis,Buchmuller:2020lbh,King:2021gmj}. The breaking of $U(1)_{B-L}$ gauge symmetry generates stable cosmic string network that can put severe bounds on model parameters. These bounds can be relaxed if the cosmic strings are metastable. The embedding of $U(1)_{B-L}$ group in $SO(10)$ GUT gauge group leads to production of metastable cosmic string network which can decay via the Schwinger production of monopole-antimonopole pairs, generating a stochastic gravitational wave background (SGWB), in the range of ongoing and future gravitational wave (GW) experiments.
%

The MSSM matter superfields reside in the $\bm{16}$ (spinorial) representation, whereas the MSSM Higgs doublet reside in $\bm{10}$ representations of $SO(10)$. The $SO(10)$ symmetry breaking to MSSM gauge group is achieved by non-zero VEV of $\bm{45}$ multiplet;
\begin{equation}
	SO(10) \xrightarrow{\langle \bm{45} \rangle} G_{\text{MSSM}} \times U(1)_{\chi} \xrightarrow{\langle \bm{16} \rangle, \langle \bm{\bar{16}} \rangle}  G_{\text{MSSM}},
\end{equation}
where $G_{\text{MSSM}} \equiv SU(3)_C \times SU(2)_L \times U(1)_Y$ is the MSSM gauge group. The $U(1)_{\chi}$ charge is defined as a linear combination of hypercharge $Y$ and $B-L$ charge,
\begin{equation}
	Q_{\chi} = Y x + Q_{B-L},
\end{equation}
with $x$ being a real constant. As a special case of $x = 0$, the model, after spontaneous breaking of $SO(10)$, can be effectively realized as $B-L$ extended MSSM, $U(1)_{\chi=B-L}$. The Higgs superfield pair ($H, \bar{H}$) belong to $(\bm{16} + \bm{\bar{16}})$ representation of $SO(10)$ and is responsible for breaking $G_{B-L}$ to MSSM. The first step gauge symmetry breaking produces magnetic monopoles which are inflated away during inflation, whereas the second step breaking produces metastable cosmic string network.

If cosmic strings form after inflation, they exhibit a scaling behavior where the stochastic GW spectrum is relatively flat as a function of the frequency, and the amplitude is proportional to the string tension $\mu_{CS}$. For our case,  $\mu_{CS}$ can be written in term of $M$ as \cite{Hill:1987ye},
\begin{equation}\label{cosmicmu}
\mu_{CS}= 2\pi M^2 y(\Upsilon), \quad y(\Upsilon) \approx \left\{ \begin{array}{ll}
1.04 \,\Upsilon^{0.195}, & \mbox{ $ \Upsilon > 10^{-2},$} \\
\frac{2.4}{\log[2/\Upsilon]}, & \mbox{ $ \Upsilon < 10^{-2}$},\end{array} \right.  
\end{equation}
where $\Upsilon = \frac{\kappa^2}{2g^2}$ with $g=0.7$ for MSSM. The CMB bound on cosmic string tension, reported by Planck 2018 \cite{Ade:2013xla,Ade:2015xua} is
\begin{eqnarray}
G \mu_{CS} \lesssim 2.4 \times 10^{-7},
\end{eqnarray}
where $G \mu_{CS}$ denotes the dimensionless string tension with the gravitational constant $G= 6.7 \times 10^{-39}~\text{GeV}^{-2}$. The observation of GWs from cosmic strings is crucially dependent on two scales; the energy scale of inflation $\Lambda_{\text{inf}}$, and the scale at which cosmic string generate the GW spectrum $\Lambda_{CS}\equiv \sqrt{\mu_{CS}}$. The amplitude of the tensor mode cosmic microwave background (CMB) anisotropy fixes the energy scale of inflation as $\Lambda_{\text{inf}}\sim V^{1/4}\sim 3.3\times 10^{16} \, r^{1/4}$ \cite{Easther:2006qu}. Using Planck 2-$\sigma$ bounds on tensor-to-scalar ratio $r$ we obtain the upper limit on scale of inflation, $\Lambda_{inf} < 1.6 \times 10^{16} $ GeV \cite{Planck:2018jri}. In our model, strings form after inflation, namely $\Lambda_{\text{inf}}>\Lambda_{CS}$, for which a stochastic gravitational wave background (SGWB) is generated from undiluted strings. The SGWB arising from meta-stable cosmic strings network are expressed relative to critical density as \cite{Blanco-Pillado:2017oxo}
\begin{align}
\Omega_\text{GW}(f) = \frac{\partial \rho_\text{gw}(f)}{\rho_c \partial \ln f}= \frac{8 \pi f (G \mu_{CS})^2}{3 H_0^2} \sum_{n = 1}^\infty C_n(f) \, P_n \,,
\label{eq:Omega}
\end{align}
where $\rho_\text{gw}$ denotes the GW energy density, $\rho_c$ is the critical energy density of the universe, and $H_0 = 100 \,h\,\textrm{km}\textrm{s}^{-1} \textrm{Mpc}^{-1}$ is the Hubble parameter. The parameter $P_n \simeq\frac{50}{\zeta(4/3)n^{4/3}}$ is the power spectrum of GWs emitted by the $n^{\rm th}$ harmonic of a cosmic string loop and $C_n(f)$ indicates the number of loops emitting GWs that are observed at a given frequency $f$
\begin{figure}[tp]
	\centering \includegraphics[width=7.90cm]{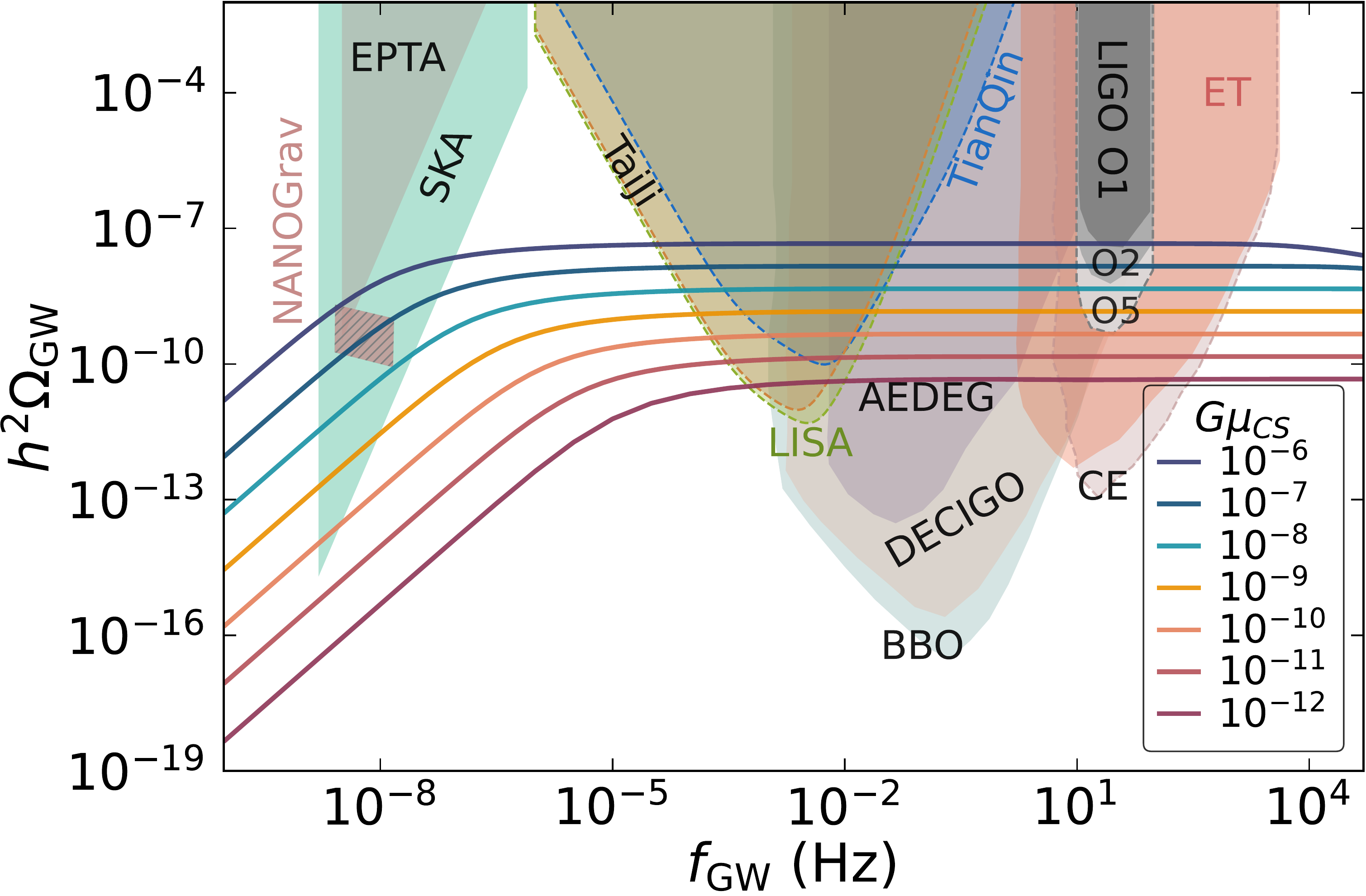}
	\centering \includegraphics[width=7.90cm]{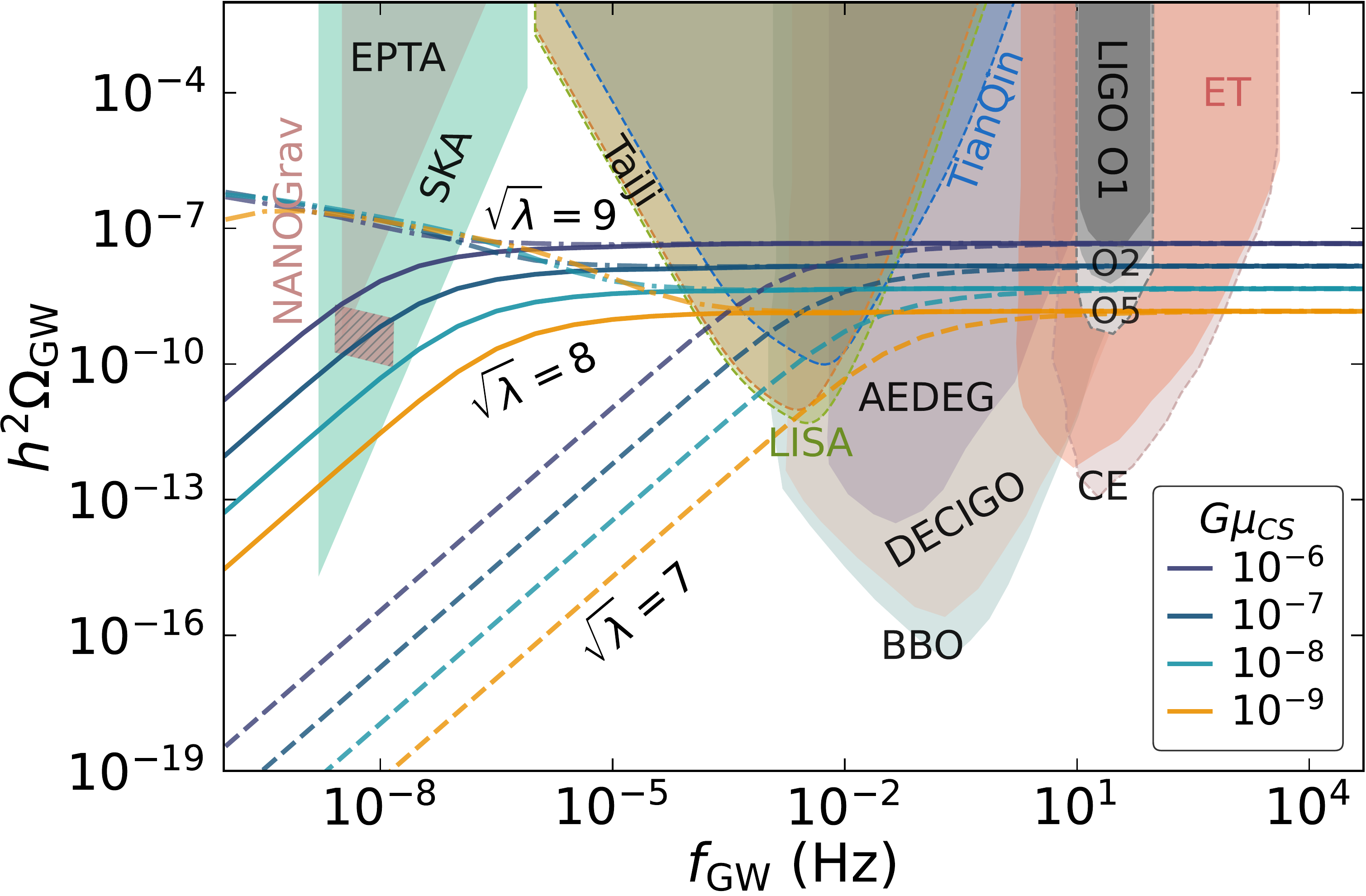}
	\caption{Gravitational wave spectra from metastable cosmic strings explaining the NANOGrav excess at 2-$\sigma$ confidence level. The curves in left panel are drawn for all range of string tension obtained in the model with $\sqrt{\lambda}=8$. The curves in right panel are drawn for the range of string tension consistent with non-thermal leptogenesis and gravitino dark matter, with $\sqrt{\lambda}=7,8,9$. The shaded areas in the background indicate the sensitivities of the current and future experiments.}
	\label{omegagw}
\end{figure}
\begin{align}
\label{eq:Cn}
C_n(f) = \frac{2 n}{f^2} \int_{z_\text{min}}^{z_\text{max}}dz\:\frac{\mathcal{N}\left(\ell\left(z\right),\,t\left(z\right)\right)}{H\left(z\right)(1 + z)^6} \,,
\end{align}
which is a function of number density of cosmic string loops $\mathcal{N}(\ell,t)$, with $\ell = 2n/((1 + z) f)$. For the number density of cosmic string loops, $\mathcal{N}(\ell,t)$, we use the approximate expressions of Blanco-Pillado-Olum-Shlaer (BOS) model given in \cite{Blanco-Pillado:2017oxo,Auclair:2019wcv}
\begin{align}
\mathcal{N}_r(\ell,t)  &= \frac{0.18}{t^{3/2}(\ell+\Gamma G\mu_{CS} t)^{5/2}},\label{eq:nr}\\
\mathcal{N}_{m,r}(\ell,t)  &= \frac{0.18\sqrt{t_{eq}}}{t^2(\ell+\Gamma G\mu_{CS} t)^{5/2}}
=\frac{0.18(2H_0\sqrt{\Omega_r})^{3/2}}{(\ell+\Gamma G\mu_{CS} t)^{5/2}}(1+z)^3~.
\end{align}
For our region of interest, the dominant contribution is obtained from the loops generated during the radiation-dominated era. For $t(z)$ and $H(z)$, we use the expressions for $\Lambda$CDM cosmology assuming a standard thermal history of universe, while ignoring the changes in the number of effective degrees of freedom with $z$ 
\begin{eqnarray}\label{OmegaGW}
H(z)&=&H_0\sqrt{\Omega_\Lambda + \Omega_m(1+z)^3+\Omega_r(1+z)^4},\\
t(z) &=& \int_{z_\text{min}}^{z_{\text{max}}} \frac{dz' }{H(z')(1+z')},\quad l(z)=\frac{2n}{(1+z)f}. 
\end{eqnarray}
The integration range in the above equation corresponds to the lifetime of the cosmic string network, from its formation at $z_\text{max} \simeq \frac{T_r}{2.7K}$ until its decay at $z_\text{min}$  given by \cite{Leblond:2009fq,Monin:2008mp,KaiSchmitz}, 
\begin{equation}
z_\text{min} = \left( \frac{70}{H_0}\right)^{1/2} \left( \Gamma  \; \Gamma_d  \; G \mu_{CS} \right)^{1/4},\quad \Gamma_d =\frac{\mu}{2\pi}e^{-\pi\lambda}, \quad \lambda = \frac{m_M^2}{\mu}
\label{zmin}
\end{equation}
where $\Gamma \simeq 50$, $m_M$ is the monopole mass, $\mu$ is the string tension, and we fix the reheat temperature at $T_r=10^8$~GeV. The dimensionless parameter $\lambda$ is the hierarchy between the GUT and $U(1)_{B-L}$ breaking scales. Fig. \ref{omegagw} shows gravitational wave spectra from metastable cosmic strings for the predicted range of cosmic string tension, $10^{-12} \lesssim G\mu_{CS} \lesssim 10^{-6}$. The curves in left panel are drawn for the GUT and the $B-L$ breaking scales ratio, $\sqrt{\lambda}=8$. The parametric space consistent with successful reheating with non-thermal leptogenesis and gravitino dark matter restricts the value of $G\mu_{CS}$ in the range $10^{-9} \lesssim G\mu_{CS} \lesssim 10^{-6}$. This is shown in the right panel of Fig. \ref{omegagw} where the curves are drawn for $\sqrt{\lambda}=7,8,9$.
It can be seen that the GW spectrum for the entire range of $G\mu_{CS}$ passes through most GW detector sensitivities.  LIGO O1 \cite{LIGOScientific:2019vic} has excluded cosmic strings formation at $G\mu_{CS} \lesssim 10^{-6}$ in the high frequency regime $10$-$100$ Hz. The low frequency band, $1$-$10$~nHz, can be probed by NANOGrav \cite{Arzoumanian:2020vkk}, EPTA \cite{Ferdman:2010xq} and other GW experiments at nano Hz frequencies. Planned pulsar timing arrays SKA \cite{Smits:2008cf}, space-based laser interferometers LISA \cite{LISA:2017pwj}, Taiji \cite{Hu:2017mde}, TianQin \cite{TianQin:2015yph}, BBO \cite{Corbin:2005ny}, DECIGO \cite{Seto:2001qf}, ground-based interferometers, such as Einstein Telescope \cite{Punturo:2010zz} (ET), Cosmic Explorer \cite{LIGOScientific:2016wof} (CE), and atomic interferometer AEDGE \cite{AEDGE:2019nxb}, will probe GW generated by metastable cosmic string in a wide regime of frequencies.

\subsection{Explaining the NANOGrav results}
\begin{figure}[tp]
	\centering \includegraphics[width=12.0cm]{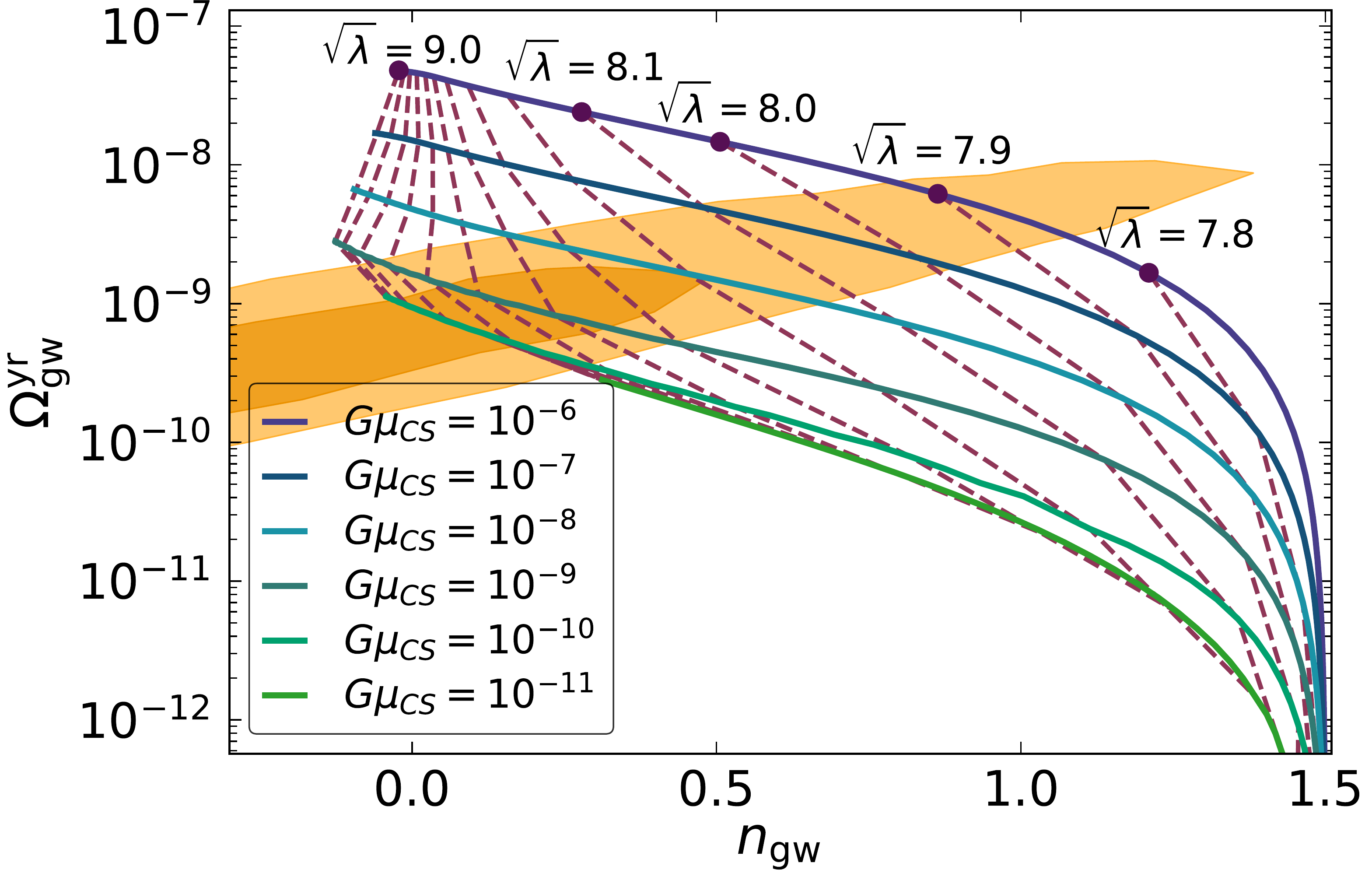}
	\caption{Gravitational wave signals from metastable cosmic strings compared to the NANOGrav observations for different values of the string tension $G \mu_{CS}$ and the hierarchy between the GUT and $U(1)$ breaking scale $\lambda$. The solid colored lines represent fixed values of $G \mu_{CS}$ whereas, the dotted lines represent contours of $\sqrt{\lambda}$. The dark (light) yellow region represent the 1-$\sigma$ (2-$\sigma$) bounds reported by NANOGrav 12.5 yr data \cite{Arzoumanian:2020vkk}.}
	\label{nano}
\end{figure}

We now discuss the SGWB signal predicted by metastable cosmic strings for recent NANOGrav 12.5 yr results \cite{Arzoumanian:2020vkk}, which constrain the amplitude and slope of a stochastic process. The amplitude of the SGWB is obtained in terms of dimensionless characteristic strain $h_c = A (f/f_\text{yr})^\alpha$ at the reference frequency $f_\text{yr}=32$~nHz as \cite{Buchmuller:2020lbh}
\begin{align}
\Omega_\text{GW}(f)  = \frac{2 \pi^2 f_\text{yr}^2 A^2 }{3 H_0^2}  \left( \frac{f}{f_\text{yr}} \right)^{2 \alpha + 2} \equiv \Omega_\text{gw}^\text{yr}\left( \frac{f}{f_{yr}} \right)^{n_{gw}}~\label{yrgw},
\end{align}
where $A$ is the strain amplitude. At low GW frequency, $\Omega_\text{GW}$ behaves as $\sim f^{3/2}$, whereas at high GW frequencies, $\Omega_\text{GW} \sim 1$. NANOGrav uses a power law fit with $5-\gamma=2+2\alpha=n_{gw}$ and constrain the parameters $A$ and $\gamma$. This allows us to directly translate the 1- and 2-$\sigma$ NANOGrav bounds given in \cite{Arzoumanian:2020vkk} into the $\Omega_\text{gw}^\text{yr}$-$n_\text{gw}$ plane, as displayed by the yellow shaded regions in Fig. \ref{nano}. Following \cite{JohnEllis}, we extract the amplitude $\Omega_\text{gw}^\text{yr}$ and slope $n_\text{gw}$ using Eq. \eqref{yrgw} by comparing the amplitude at the pivot scale $f_*$ and taking the logarithmic derivative of $\Omega_\text{gw}(f)$ at the desired frequency scale $f_*$ \footnote{Here we have employed numerical differentiation method. For least squares power-law fit method, see \cite{Buchmuller:2020lbh}},
\begin{eqnarray}
n_\text{gw}&=& \left.\frac{d\log{\Omega_\text{GW}(f)}}{d\log{f}}\right|_{f=f_*},\\
\Omega_\text{gw}^\text{yr}&=&\Omega_\text{GW}(f_*)\left( \frac{f_\text{yr}}{f_*}\right)^{n_\text{gw}}.
\end{eqnarray}
Fig. \ref{nano} shows the comparison of the predictions from metastable cosmic strings (mesh of solid and dotted curves) with the constraints on the amplitude and tilt from \cite{Arzoumanian:2020vkk} (yellow shaded region). We vary $G \mu_{CS}$ from $10^{-11}$ to $10^{-6}$, however the CMB constraint $G\mu_{CS}\leq 1.3\times 10^{-7}$ only applies to cosmic strings with a life-time exceeding CMB decoupling, corresponding to $\sqrt{\lambda} \gtrsim 8.6$. For each value of $G\mu_{CS}$, we consider the GUT and the $B-L$ breaking scales ratio in the range $\sqrt{\lambda} = 7.4-9.0$, where smaller values lead to a small spectrum at nHz frequencies that can be detected by future experiments, while all values $\sqrt{\lambda} \gtrsim 8.8$ quickly converge towards the result for stable cosmic strings and can be observed by NANOGrav and PPTA experiments \cite{Arzoumanian:2020vkk}. The parametric space in the above model, consistent with successful reheating with non-thermal leptogenesis and gravitino dark matter restrict the allowed values of string tension to the range $10^{-9} \lesssim G\mu_{CS} \lesssim 8 \times 10^{-6}$ that lies within the 1- and 2-$\sigma$ bounds of NANOGrav, as well as the sensitivity bounds of future gravitational wave (GW) experiments. 

\section{Summary}
To summarize, we have investigated various cosmological implications of a generic model based on the $U(1)_{B-L}$ extension of the MSSM gauge symmetry in a no-scale K\"ahler potential setup, highlighting the issues of inflation, leptogenesis and baryogenesis, gravitino as well as the stochastic gravitational wave background (SGWB) from metastable cosmic sting network. The embedding of $U(1)_{B-L}$ into a simply-connected group $SO(10)$, produces metastable cosmic string due to the spontaneous pair creation of a monopole and an anti-monopole, which can generate a stochastic gravitational wave background (SGWB) in the range of ongoing and future gravitational wave (GW) experiments. The interaction between $U(1)_{B-L}$ Higgs and the neutrino superfields generate heavy Majorana masses for the right-handed neutrinos. The heavy Majorana masses explain the tiny neutrino masses via the seesaw mechanism, a realistic scenario for reheating and non-thermal leptogenesis. A wide range of reheat temperature $(10^{7} \lesssim T_r \lesssim 10^{10}) ~ \text{GeV}$ and $U(1)_{B-L}$ symmetry breaking scale $(1.3 \times 10^{13} \lesssim M \lesssim 2.0 \times 10^{16}) ~ \text{GeV}$ is achieved here with successful non-thermal leptogenesis and stable gravitino as a possible dark matter candidate. The metastable cosmic string network admits string tension values in the range $10^{-12} \lesssim G\mu_{CS}\lesssim 8 \times 10^{-6}$. A successful reheating with non-thermal leptogenesis and gravitino dark matter restrict the allowed values of string tension to the range $10^{-9} \lesssim G\mu_{CS} \lesssim 8 \times 10^{-6}$, predicting a stochastic gravitational-wave background that lies within the 1-$\sigma$ bounds of the recent NANOGrav 12.5-yr data, as well as within the sensitivity bounds of future GW experiments.

\section*{Acknowledgments}
 We thank Valerie Domcke, George K Lenotaris and Kazunori Kohri for valuable discussions.The work of S.N is supported by the United Arab Emirates University under UPAR Grant No. 12S004.


\end{document}